\documentclass[twocolumn,twocolappendix]{aastex631}

\begin{document}

\title{Impact of Electron Precipitation on Brown Dwarf Atmospheres and the Missing Auroral H$_{3}^{+}$ Emission}


\correspondingauthor{J.\ Sebastian Pineda}
\email{sebastian.pineda@lasp.colorado.edu}

\author[0000-0002-4489-0135]{J. Sebastian Pineda}
\affiliation{University of Colorado Boulder, Laboratory for Atmospheric and Space Physics, 3665 Discovery Drive, Boulder, CO 80303, USA}

\author{Gregg Hallinan}
\affiliation{California Institute of Technology, Department of Astronomy, 1200 E. California Ave, Pasadena CA, 91125, USA}

\author{Jean Michel Desert}
\affiliation{Anton Pannekoek Institute, University of Amsterdam, Amsterdam, 1098XH, Netherlands}

\author{Leon K.\ Harding}
\affiliation{Northrop Grumman Space Systems, Redondo Beach, CA 90278, USA}

\begin{abstract}

Recent observations have demonstrated that very-low mass stars and brown dwarfs are capable of sustaining strong magnetic fields despite their cool and neutral atmospheres. These kG field strengths are inferred based on strong highly circularly polarized GHz radio emission, a consequence of the electron cyclotron maser instability. Crucially, these observations imply the existence of energetic non-thermal electron populations, associated with strong current systems, as are found in the auroral regions of the magnetized planets of the Solar System. Intense auroral electron precipitation will lead to electron collisions with the H$_{2}$ gas that should generate the ion H$_{3}^{+}$. With this motivation, we targeted a sample of ultracool dwarfs, known to exhibit signatures associated with aurorae, in search of the K-band emission features of H$_{3}^{+}$ using the Keck telescopes on Mauna Kea. From our sample of 9 objects, we found no clear indication of H$_{3}^{+}$ emission features in our low-medium resolution spectra (R$\sim$3600). We also modeled the impact of an auroral electron beam on a brown dwarf atmosphere, determining the depth at which energetic beams deposit their energy and drive particle impact ionization. We find that the H$_{3}^{+}$ non-detections can be explained by electron beams of typical energies $\gtrsim$2-10~keV, which penetrate deeply enough that any H$_{3}^{+}$ produced is chemically destroyed before radiating energy through its infrared transitions. Strong electron beams could further explain the lack of UV detections, and suggest that most or nearly all of the precipitating auroral energy must ultimately emerge as thermal emissions deep in brown dwarf atmospheres. 

\end{abstract}


\section{Introduction}

The discovery of strong radio emission from brown dwarfs \citep[e.g.,][]{Berger2001,Hallinan2007,Hallinan2008,Berger2009,Route2012, Burgasser2015b,Kao2016} has opened up a new window into the magnetic activity of ultracool dwarfs (UCDs; spectral type $\ge$ M7). Whereas typical tracers of stellar activity, like H$\alpha$ and X-ray, appear to decline into the UCD regime \citep[see within][]{Pineda2017}, such objects with radio emission appear to display consistent flux levels from late M dwarfs to T dwarfs, allowing for continued investigation into the magnetic fields and dynamos operating in brown dwarfs \citep{Kao2016, Route2016, Kao2018ApJS..237...25K}. Furthermore, detailed study of individual objects such as LSR~J1835+3259 and TVLM513-46546 has revealed the importance of the electron cyclotron maser instability (ECMI) as the key mechanism generating the highly polarized coherent strong periodic radio pulses \citep{Hallinan2008, Lynch2015, Hallinan2015, Williams2015a}. Concurrently, multiple studies have also observed these objects displaying periodic optical variability consistent with the rotational period seen in the radio pulses \citep[e.g.,][]{Littlefair2008}, with \cite{Harding2013} suggesting a possible correlation between the optical variability and the radio emission.
 
To test this potential connection, \cite{Hallinan2015} observed LSR~J1835+3259 simultaneously at radio and optical wavelengths, demonstrating that the object's H$\alpha$ emission was periodic and consistent with the rotational period determined from the radio observations. Through spectroscopic monitoring they also demonstrated that the broadband optical variability was connected with the H$\alpha$ emission and generated both correlated and anti-correlated light curves in broadband photometry, depending on the wavelengths of the observations. They concluded that the electron beam responsible for the radio emission impacted the atmosphere and created the surface feature seen both in H$\alpha$ and broadband optical monitoring, analogous to what is observed in the auroral emissions of the magnetized planets of the Solar System \citep{Hallinan2015}.
 
This analogy suggests that brown dwarfs are capable of hosting auroral phenomena in their hydrogen dominated atmospheres similar to what is seen on Jupiter and Saturn. In these gas giant planets, the ECMI radio emission is generated by a strong magnetospheric current system that produces an energetic non-thermal electron beam that precipitates into the atmosphere depositing the bulk of its energy \citep[see][and references therein]{Keiling2012,Badman2015}. This atmospheric interaction generates a cascade of auroral emission processes at ultraviolet, optical, and infrared (IR) wavelengths \citep[see][and references therein]{Badman2015}. 

Of particular importance is the collisional ionization of molecular hydrogen and the subsequent production of H$_{3}^{+}$, through the chemical reaction 

\begin{equation}
\mathrm{H}^{+}_{2} \; + \; \mathrm{H}_{2} \rightarrow \; \mathrm{H}^{+}_{3} + \mathrm{H} \; .
\end{equation}

\noindent In auroral atmospheric regions, this reaction proceeds very rapidly creating large quantities of the triatomic hydrogen ion from the impact of the electron beam. H$_{3}^{+}$ does not have an electric dipole moment but instead emits strongly from the rotational-vibrational modes of its fundamental band at 4 $\mu$m and in the overtone band at 2 $\mu$m \citep{Maillard2011}. The spectral lines comprise a forest of discrete and narrow features in each band. This ion of hydrogen plays a significant role in regulating the exospheric temperature, as one of the most strongly emitting species in the Jovian ionosphere \citep{Maillard2011}. Moreover, the long timescale and the thermal nature of the transitions means that the H$_{3}^{+}$ emission is sensitive to the properties of the atmosphere \citep{Tao2011, Badman2015}. Indeed, \cite{Tao2012} showed how the properties of the electron beam and the background atmospheric temperature can be determined from a study of the H$_{3}^{+}$ emission lines of Jupiter.

If brown dwarfs can generate auroral emissions, as suggested by the ECM radio emission, then they become strong candidates for the search for H$_{3}^{+}$ emission in atmospheres beyond the Solar System. These efforts have focused on known hot jupiters \citep[e.g.,][]{Shkolnik2006,Lenz2016}, where the intense radiation and close proximity to their host stars could strengthen the signatures of auroral processes as the planet interacts with the star's stellar wind and intense high energy radiation \citep[e.g.,][]{Zarka2007,Chadney2016A&A...587A..87C}. However, these efforts have not yet yielded any detections \citep{Shkolnik2006,Lenz2016,Gibbs2022AJ....164...63G}. 

Brown dwarfs, on the other hand, could be potentially more favorable targets for the detection of H$_{3}^{+}$. Unlike the hot jupiters, isolated brown dwarfs can be studied directly without having to remove the effect of a bright stellar host. Moreover, based on the observed radio emissions, the strength of the auroral emission in brown dwarfs can be several orders of magnitude greater than what is seen in the Solar System, because of their rapid rotation rates and strong kG magnetic field strengths \citep{Hallinan2015}. The warmer temperatures of the brown dwarf atmospheres may also push more of the prominent H$_{3}^{+}$ auroral thermal energy to emerge in the overtone band at 2~$\mu$m. Because of the similar atmospheric temperatures, the potential detection of H$_{3}^{+}$ in brown dwarf atmospheres could also be used as a stepping stone toward the exploration of H$_{3}^{+}$ in hot jupiter atmospheres and an understanding of the physics underlying auroral processes in an entirely new parameter space. Recently, \citet{Gibbs2022AJ....164...63G} searched a sample of brown dwarfs for H$_{3}^{+}$ emissions at 4~$\mu$m with Keck NIRSPEC, reporting upper limits on the four auroral targets included in their sample.

With this work, we report a similar search across a broader sample of 9 targets (some overlapping with \citealt{Gibbs2022AJ....164...63G}), focusing on the overtone band at 2~$\mu$m, K-band, which is readily accessible from the ground and has recently seen improved sensitivity with updated IR instruments. In Section~\ref{sec:data}, we review our targets, observing strategy, and data reduction. In Section~\ref{sec:results}, we present the results of our observational survey. In Section~\ref{sec:ebeam}, we determine at which height in the atmosphere we may expect brown dwarf aurural energy deposition. Lastly, in Section~\ref{sec:discuss}, we discuss our findings and the implications for studies of auroral phenomena in the UCD regime.

\section{Observations and Data}\label{sec:data}

We selected a sample of ultracool dwarfs with known indicators of auroral activity to search for H$_{3}^{+}$ emissions. These targets are listed in Table~\ref{tab:obslog}, and we provide detailed information for each one in Appendix~\ref{sec:targ}, including physical properties and notes of interest.

\begin{figure}[htbp]
	\centering
	\includegraphics{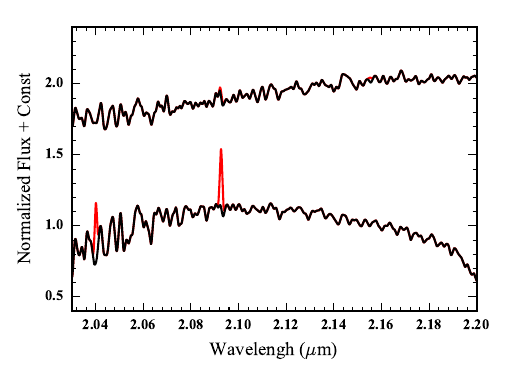} 
	\caption{PHOENIX model spectra (black) of $\log$ g = 5 (cgs) and effective temperatures 2300 K and 1200 K, top line and lower line respectively, with H$_{3}^{+}$ emission lines superimposed (red). The models are representative of the typical stars in our target sample (see Section~\ref{sec:targ}) and indicate what the spectra would look like if the prominent H$_{3}^{+}$ emission lines carried as much energy as is found in H$\alpha$ emission ($L_{\mathrm{H}\alpha}/L_{\mathrm{bol}} \sim10^{-5.5}$; see Section~\ref{sec:scale}). The spectra are normalized by the mean flux between 2.1 and 2.2 $\mu$m. The emission is more difficult to detect in the warmer objects which have more photospheric flux in the K-band than in the cooler brown dwarfs.}
	\label{fig:simLine}
\end{figure}

\subsection{H$_{3}^{+}$ K-band Energy}\label{sec:scale}

The detection of auroral radio emission in brown dwarf atmospheres indicates that auroral processes may be creating significant quantities of H$_{3}^{+}$ in their upper atmospheric regions. Using the relative energy scaling observed in the Jovian atmosphere between the different multi-wavelength auroral emissions, we can estimate the amount of energy in H$_{3}^{+}$ K-band emission features we expect to see from UCDs. In Jupiter, about a $\sim$10\% of the total auroral energy emerges in H$_{3}^{+}$ emission lines, with the total emission around 4 $\mu$m comprising $\sim$85\%-90\% of the H$_{3}^{+}$ energy \citep{Bhardwaj2000}. However, compared to the energy in H$\alpha$ ($\lesssim$1\% of total) and radio ($\lesssim$0.1\% of total), the H$_{3}^{+}$ features carry considerably more energy.

\begin{figure}[htbp]
	\centering
	\includegraphics[width=0.45\textwidth]{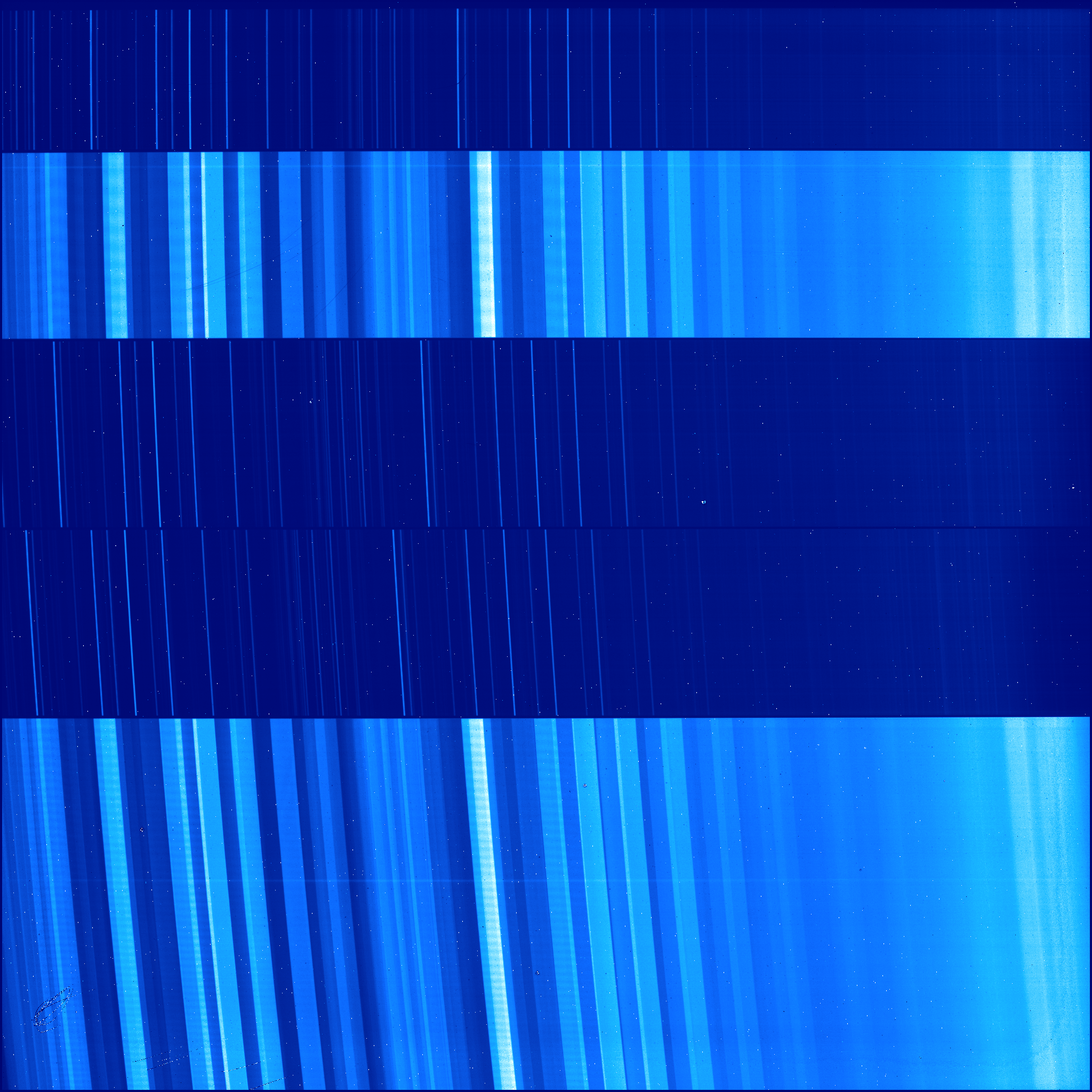} 
	\caption{A raw MOSFIRE exposure from the observing night of 2014 June 14, targeting J1254-01. The raw frame illustrates the combination of narrow slits and wide slots on the CSU which enable multi-functional observing sequences by dithering between an open slit, for spectrophotometry and in order to limit the effects of narrow sky emission features, and a narrow slit, for precise control of spectral resolution. Bright dots in the data correspond to hot pixels.}
	\label{fig:mosraw}
\end{figure}

If we assume that all of the visible auroral energy on Jupiter ($\sim$100 GW) emerges as H$\alpha$ and that only 10\% of the H$_{3}^{+}$ energy emerges in K-band ($\sim$600 GW), then the expected energy in the overtone band is at least several factors ($\sim$6) greater than the H$\alpha$ energy. However, the H$\alpha$ emission is only a portion of the visible aurorae, additional features of H$_{2}$ and broadband variations likely contribute to the total visible flux \citep{Ingersoll1998}. Thus, the K-band H$_{3}^{+}$ lines likely carry at least as much energy as H$\alpha$ and are possibly several factors stronger. The clear H$\alpha$ detections of our target objects suggests that the H$_{3}^{+}$ emission features could be strong enough to detect despite the thermal photospheric emission in K-band. 

We show an example of H$_{3}^{+}$ emission features with energy equivalent to that of H$\alpha$, $L_{\mathrm{H}\alpha}/L_{\mathrm{bol}} \sim 10^{-5.5}$, in Figure~\ref{fig:simLine}, sitting on top of the K-band photospheric emission represented by PHOENIX BT-Settl model spectra \citep{Baraffe2015}. The figure shows two representative spectra normalized by the mean flux between 2.1 and 2.2 $\mu$m and offset by a constant, the top having T$_{\mathit{eff}}$ = 2300 K and $\log$ g = 5 and the lower line having T$_{\mathit{eff}}$ = 1200 K and $\log$ g = 5. Because of the bright photospheric flux in the K-band the emission features are harder to distinguish for the warmer object than they are for the cooler object in which the emission stands out clearly. Similar calculations can be done using the radio emission, however, any auroral H$\alpha$ is physically associated with the atmospheric conditions producing the H$_{3}^{+}$ and is more likely to yield an accurate estimate. 

Although UCDs have been observed extensively at IR wavelengths in the past, these observations have predominantly been at very-low resolution ($\sim$~500), or have not focused on potentially auroral brown dwarf targets with deep K-band observations, precluding the successful identification of these features \citep[e.g.,][]{McLean2003,Rayner2009, Blake2010}. Furthermore, the H$_{3}^{+}$ emission features are likely to be very narrow, requiring highly sensitive and at least medium resolution spectrographs. The deep observations are essential, especially for late M dwarfs and early L dwarfs, in order to distinguish from the photospheric emission. Although this is not an issue for the cooler objects, the faint K-band magnitudes of the T dwarfs makes high resolution observations difficult. The development of highly sensitive IR instruments provides the opportunity to look for these features in brown dwarf atmospheres.

\begin{deluxetable*}{l c c c c c c c c}
	\tablecaption{Targets and Observations for our H$_{3}^{+}$ MOSFIRE Survey \label{tab:obslog}
	} 
	\tablehead{ \colhead{Target} & \colhead{Spectral Type} &   \colhead{UT Date} &  \colhead{t$_{\mathrm{exp}}$ (s)} & Total Time (s) &\colhead{Selection} &\colhead{Seeing}  & \colhead{Slit width}  & \colhead{Reference} }
	\startdata
	LSR J1835+3259  & M8.5   &2013 March 30& 23 &  186 & H$\alpha$/ Radio &  1".0 & 0''.5 & 7  \\
	TVLM513-46546 &	M9       &2013 March 30&  23 &186  &	H$\alpha$/Radio & 1".0 & 0''.5 & 8 \\
	2MASS J0746+2000  & L0+L1.5 &2013 October 11 & 25 & 1508    &	H$\alpha$/Radio &0''.9 & 0''.5 & 4 \\ 
	DENIS J1058-1548  &	L3 &2014 March 17& 29 &1513  & H$\alpha$/Variable\tablenotemark{a} &0''.8 & 0''.7& 3 \\ 
	2MASS J0036+1821  &L3.5 & 2013 October 12 & 25 &2910 & H$\alpha$/Radio &  0''.8 & 0''.3 & 4, 6 \\
	2MASS J1254-0122 &T2 & 2013 June 14 & 29 &1513 & H$\alpha$/Variable\tablenotemark{a} & 0''.4 & 12'' & 5\\ 
	SIMP J0136+0933 & T2.5 & 2013 October 12& 29 &1047  &	Radio/Variable\tablenotemark{a} & 0''.9 & 0''.5 & 1 \\
	2MASS J1237+6526 &T6.5& 2014 March 17 & 60 & 1192 & H$\alpha$/Radio& 0''.8 & 0''.7  & 2 \\
	2MASS J1047+2124 &T6.5&2013 March 30&  119 & 1432 & Weak H$\alpha$/ Radio & 1''.0 & 1''.0 & 1 \\
	\enddata
	\tablenotetext{a}{ Variable targets are objects with known near infrared photometric variability.}
	\tablenotetext{}{R\textsc{eferences.} -- (1) \citealt{Artigau2006}, (2) \citealt{Burgasser1999}, (3) \citealt{Delfosse1997}, (4) \citealt{Kirkpatrick2000}, (5) \citealt{Leggett2000}, (6) \citealt{Reid2000}, (7) \citealt{Reid2003}, (8) \citealt{Tinney1993}. }
\end{deluxetable*}

\begin{figure}[htbp]
	\centering
	\includegraphics{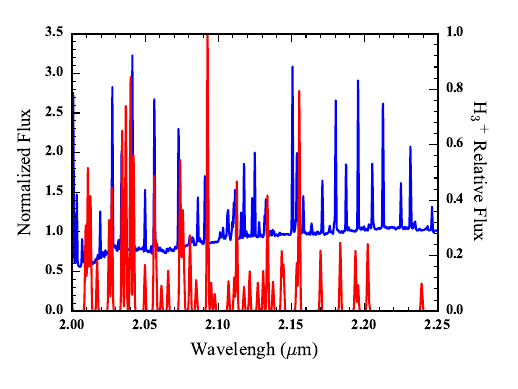} 
	\caption{The error spectrum (left axis -- blue line), normalized to its median level, of our observations of DENI1058-15 (R$\sim$3600), illustrating the locations of the sky emission lines as narrow peaks in the uncertainty with a relative flux spectrum (right axis - red line) of H$_{3}^{+}$ overlaid (T = 1800 K), showing the locations of the K-band emission features (convolved to instrument resolution). The H$_{3}^{+}$ spectrum's most prominent feature is the emission line at $\sim$2.09 $\mu$m which falls between sky emission lines.}
	\label{fig:DENIS_sky}
\end{figure}

\subsection{Observational Data}

In order to search for H$_{3}^{+}$ emission features, we focused on the 2$\nu_{2}$ overtone band around 2 $\mu$m in the K-band, where we expect potentially strong H$_{3}^{+}$ emission features. We observed our target sample (see Table~\ref{tab:obslog}) using the Mulit-Object Spectrometer For Infrared Exploration (MOSFIRE) on the Keck Telescopes on Mauna Kea \citep{McLean2010,McLean2012}. MOSFIRE is a multi object spectrograph with a configurable slitmask unit (CSU) with movable bars that can be repositioned as desired during the observations. MOSFIRE uses a highly sensitive teledyne H2RG HgCdTe detector with 2k x 2k pixels within a 6'.1 x 6'.1 field of view, and is designed to provide NIR spectra of faint objects (usually galaxies), but has been used for careful spectrophotometric work during exoplanet transits \citep[e.g.,][]{Crossfield2013}. MOSFIRE provides a resolution of R$\sim$3600 in the K-band with a nominal 0''.7 wide slit, spanning 1.95-2.4 $\mu$m. The resolution in each of our observations varied according to the slit width we used, from $\sim$2500 to $\sim$8400 (see Table~\ref{tab:obslog} and below for discussion).

Our observations took place over the course of 2013 and 2014 (see Table~\ref{tab:obslog}), usually only one or two objects on a given night with total exposure times ranging from 3 minutes, for the two bright M dwarfs in our sample, and up to 50 minutes for one of our targets, but usually only about 25 minutes. Our program was clouded out on dedicated observing nights, however we were able to take advantage of cooperation with companion programs looking for photometric variability in brown dwarfs and/or exoplanet transmission spectroscopy. During periods between transits and/or when specific targets were not yet sufficiently high in the sky we were able to collect deep K-band spectra of our sample target list. Consequently, the time available on each target varied considerably and we were not always able to acquire telluric calibrators to correct the K-band spectra. However, the prominent H$_{3}^{+}$ emission features avoid the regions of intense telluric absorption in the K-band around 2.02 and 2.07 $\mu$m. We present these spectra in Section~\ref{sec:results}.

We configured the CSU to contain additional reference targets to provide a check on any astrophysical variations we observed and included both narrow slit and wide slit sections of the slit mask. These can been seen in the raw exposure example, shown in Figure~\ref{fig:mosraw}. The bright bands in the image correspond to the sky emission lines seen through the wide open slit (width $\sim$10'') and can be traced vertically to the sections with narrow sky lines from the narrow slit width ($\sim$0''.7). During the observations we were able to dither along the slit direction on the mask to change between one mode and the other. This proved more expedient than moving the CSU bars, because it was quicker and slight differences between the repositioning of the CSU bars can introduce systematics into the observations from frame to frame \citep{Crossfield2013}. Although there can be differences introduced from the dithering due to slight shifts in the pointing potentially shifting the target slightly within the wide slits, these proved inconsequential for the search for H$_{3}^{+}$ emission.

The use of both slit widths allowed us to provide precise control of the observed spectral resolution with the narrow slits while using the wide slits for spectrophotometry and to help distinguish between sky lines and narrow H$_{3}^{+}$ emission features. Although the the wide slits increase the background noise by spreading out the bright night sky lines, it eliminates any issues dealing with residual sky line subtraction and confusing a poorly subtracted narrow skyline with potential astrophysical emission features. The sky emission varies on the order of 1-2 minutes and we found that for careful sky line subtraction, we required exposures of $\leq$30 s (see Table~\ref{tab:obslog}). Nevertheless, we were able to use our wide slit exposures as a verification when assessing any narrow features in the observed spectra. Additionally, the most prominent H$_{3}^{+}$ features avoid the major OH sky lines. We illustrate this in Figure~\ref{fig:DENIS_sky} showing the error spectrum from our observations of DENIS1058-15 with an H$_{3}^{+}$ spectrum overlaid. The spikes in the error spectrum correspond to the locations of sky emission lines, with the strongest H$_{3}^{+}$ lines usually located between sky features. In Section~\ref{sec:results}, we present our best co-added spectra (based on S/N and resolution) for each object, usually from narrow slit observations. Our observations of J1254-01 were conducted under good seeing conditions (see Table~\ref{tab:obslog}) and we thus present the wide slit observations of that target because of their high resolution and high signal-to-noise.

\begin{figure*}[htbp]
	\centering
	\includegraphics{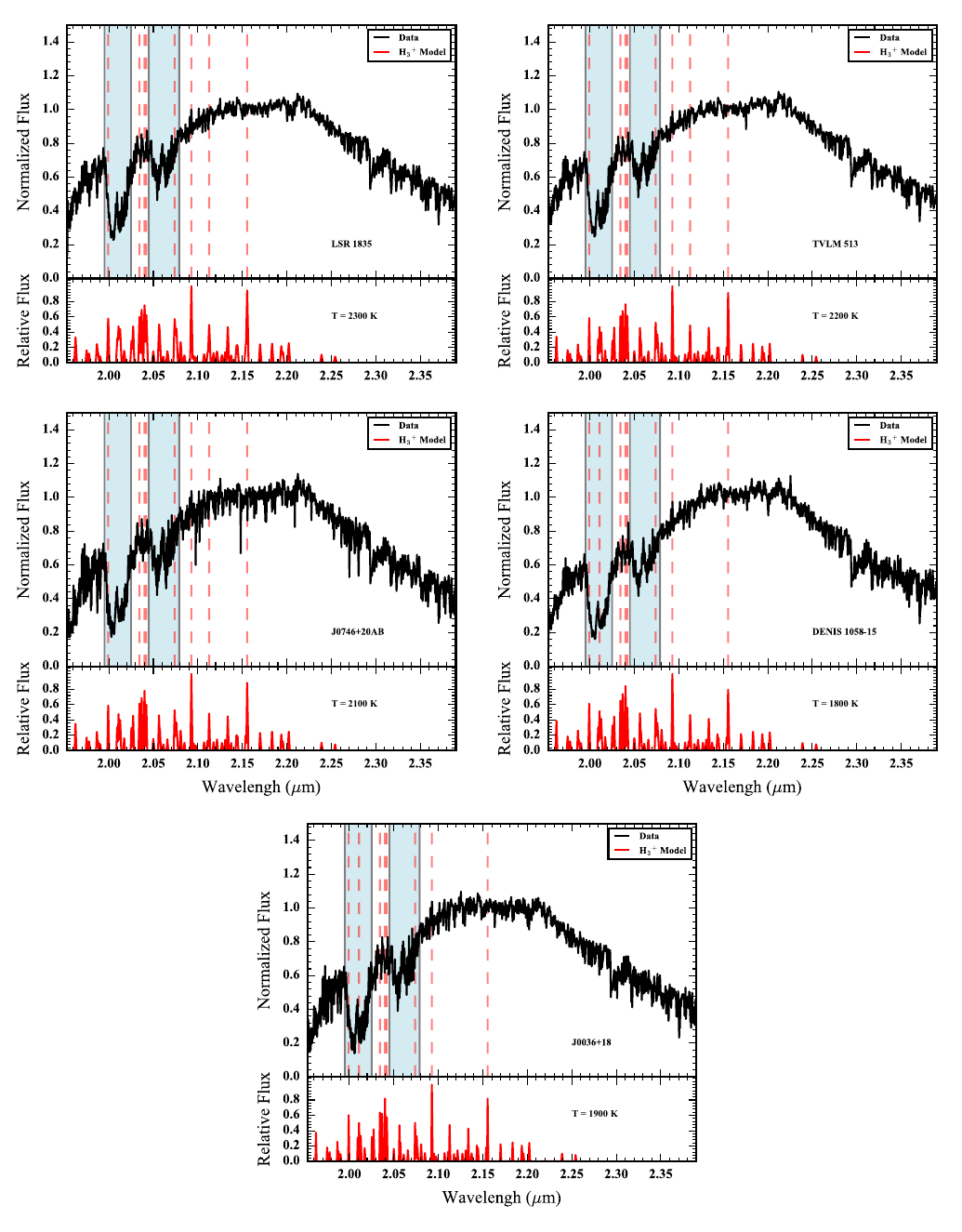} 
	\caption{The K-band MOSFIRE spectrum of LSR 1835 (top left), TVLM 513 (top right), J0746+20AB (center left), DENIS 1058-15 (center right), and J0036+18 (bottom) in black, normalized to the median flux level between 2.1 and 2.2 $\mu$m. The lower panel of each plot shows a simulated spectrum of H$_{3}^{+}$, with a rotational and vibrational temperature of 2300 K (top left), 2200 K (top right), 2100 K (center right), 1800 K (center left), and 1900 K (bottom). The shaded regions denote significant telluric absorption and the dashed lines trace the locations of the eight strongest lines of H$_{3}^{+}$, in the respective plots.}
	\label{fig:MOScomp1}
\end{figure*}

\begin{figure*}[htbp]
	\centering
	\includegraphics{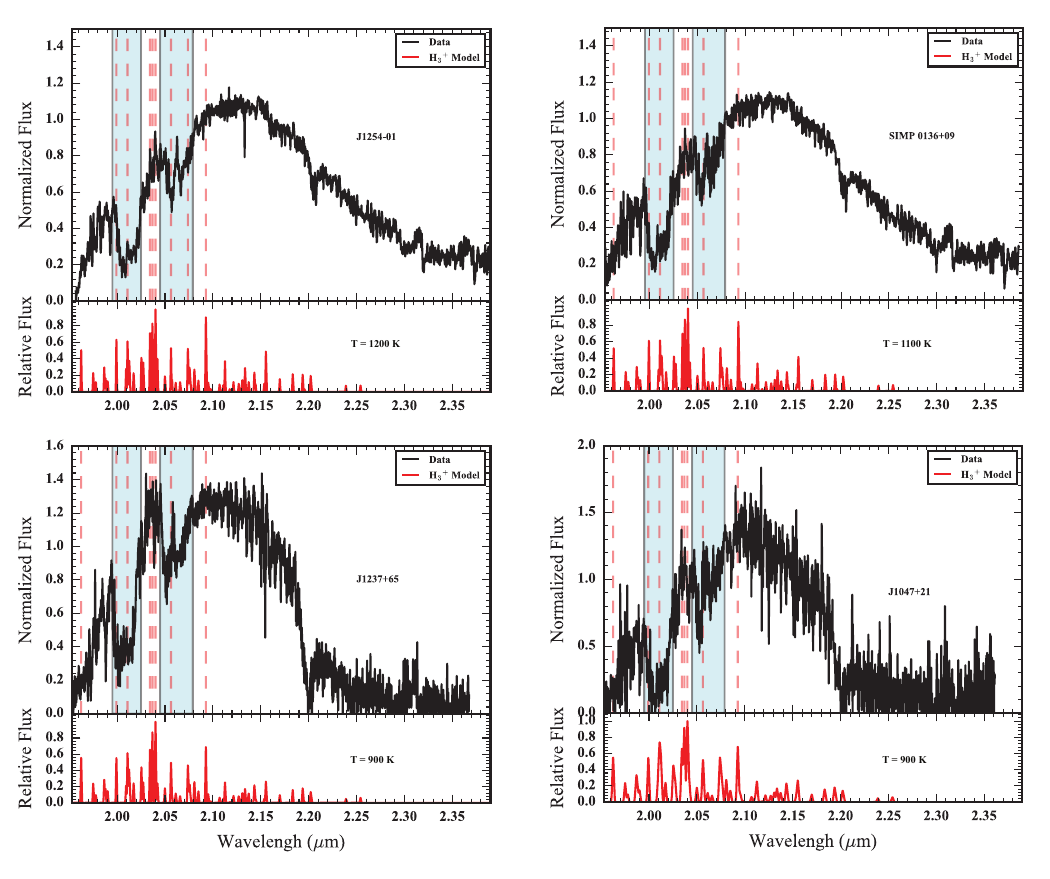} 
	\caption{Same as Figure~\ref{fig:MOScomp1}, but for J1254-01 (top left), SIMP0136+09 (top right), J1237+65 (bottom left), and J1047+21 (bottom right). The lower panel of each plot shows a simulated spectrum of H$_{3}^{+}$, with a rotational and vibrational temperatures of 1200 K (top left), 1100 K (top right), and 900 K (bottom).}
	\label{fig:MOScomp2}
\end{figure*}

For each target we used an ABBA nodding pattern, to allow clean background subtraction between adjacent frames with typical nodding sizes of at least 4''.  The IR detector was readout  in the multiple correlated double sampling mode (MCDS), 4, 8 or 16, depending on the exposure time, available observing time and the brightness of the object. While additional readouts help limit the read noise in each exposure, it adds additional overhead which may limit the total signal to noise that is ultimately achieved over the course of many exposures within a given amount of time. For the brighter objects, we typically used MCDS 4 and MCDS 16 for the fainter objects with longer exposures.

The data were reduced using the MOSFIRE Data Reduction Pipeline (DRP)\footnote{\url{https://keck-datareductionpipelines.github.io/MosfireDRP/}} to perform the flat fielding, background subtraction through frame differencing, cosmic ray rejection, co-addition, and wavelength calibration. We took flat fields and any necessary arc lamps at the beginning and end of the night with the CSU configured to our science masks, and when possible prevented changes to the CSU between calibration and science exposures to limit systematic effects. The flats in K-band require the use of dome lamp exposures and additional `lamp off' exposures to assess the thermal background contribution to the `lamp-on' signal. Additionally, because the lamps were too bright  to properly calibrate the wide slits, we took many additional ($\sim$50) lamp-off exposures to construct accurate flat fields with the wide slits.\footnote{Since these observations, Keck Observatory has provided variable power lamps to more easily use wide slit observing modes see \url{http://www2.keck.hawaii.edu/inst/mosfire/}.} In the K-band, the OH sky lines are used for the primary wavelength calibration, however they become scarce at the long wavelength end of the band and are thus supplemented by arc lamp spectra with the solutions being merged and corrected for the lamp optical path differing from the astrophysical signal path. The DRP produces 2D spectra from which we use optimal extraction to create our reduced science spectra.

\section{Results}\label{sec:results}

We report the results of our survey here, plotting each of the spectra from our MOSFIRE observations in Figures~\ref{fig:MOScomp1}-\ref{fig:MOScomp2}. In each of panel of these plots, we show in the top section our K-band spectrum and in the lower section an H$_{3}^{+}$ spectrum. The shaded region marks off where there is significant telluric absorption and the dashed lines show the location in the top panel of the eight highest peaks of the H$_{3}^{+}$ spectrum below. The expected H$_{3}^{+}$ spectrum is taken from the intensity calculator of the H$_{3}^{+}$ Resource Center.\footnote{\url{http://h3plus.uiuc.edu/}} The calculator uses input rotational and vibrational temperatures and a computed line list to construct the intensity spectra following \cite{Neale1996}. We use equivalent rotational and vibrational temperatures consistent with the photospheric temperatures of the UCDs (see Section~\ref{sec:targ}). Equating the two temperatures, assumes the molecule has thermalized in the atmosphere, which is consistent with the Jovian observations \cite[e.g.,][]{Miller1990}. Although the exospheric temperature is likely warmer, as is seen in Jupiter, there can be large variations and the photosphere presents a representative value for the atmosphere. The shape of the spectrum and the prominent lines do not change considerably with changes in the temperature; this is evident by comparing the H$_{3}^{+}$ spectra plotted for each of the targets with different temperatures spanning 800-2300 K.

In these spectra, for all but the faint T dwarf targets, J1237+65 and J1047+21, which have lower signal-to-noise observations, the variations reflect the the physical features shaping the K-band flux of these objects --- the jagged appearance of the data in these plots is not noise but the collection of closely spaced absorption lines.

\begin{figure*}[htbp]
	\centering
	\includegraphics{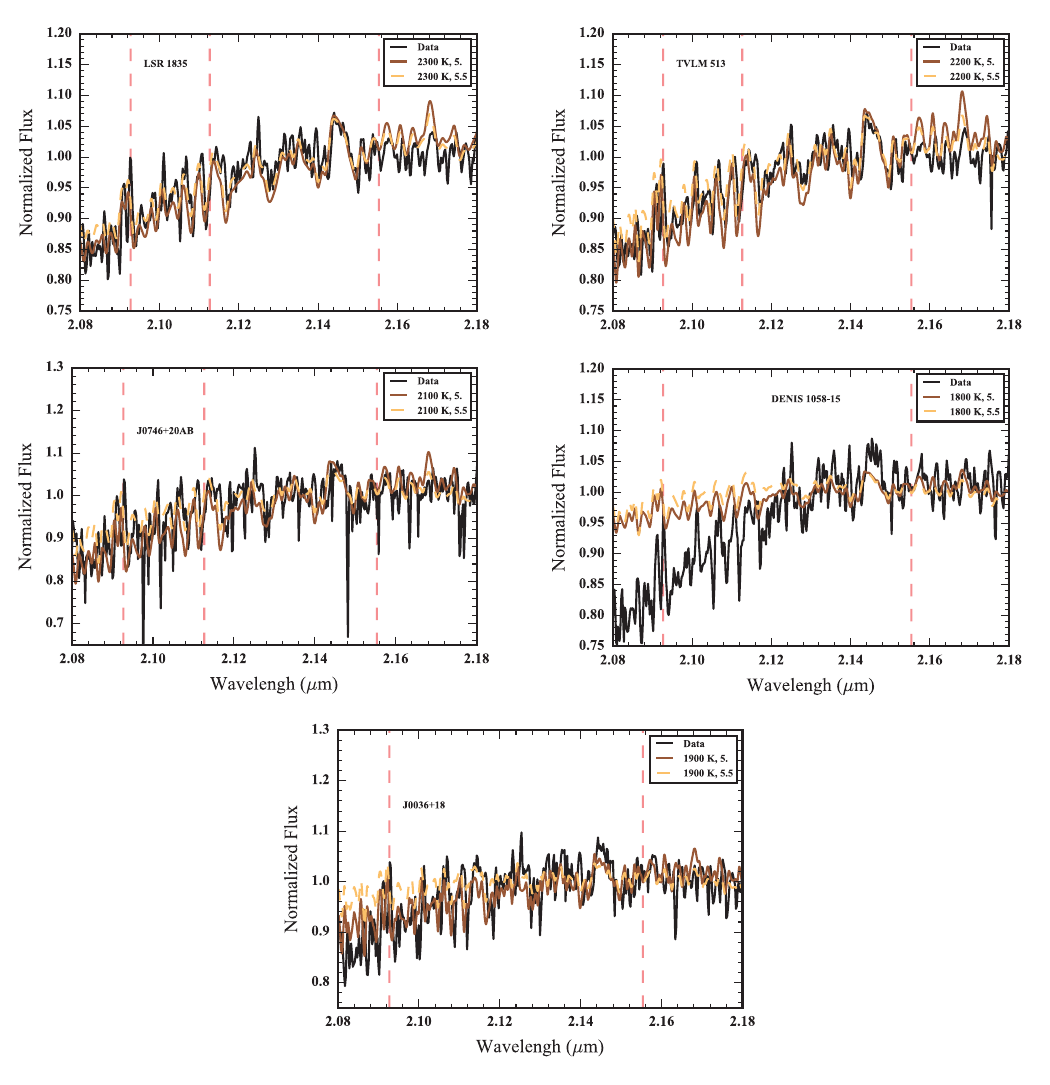} 
	\caption{A close in view of our MOSFIRE spectrum (in black) highlighting the locations of the strongest expected H$_{3}^{+}$ features with PHOENIX model spectra overlaid for LSR 1835 (top left), TVLM~513 (top right), J0746+20AB (center left), DENIS~1058-15 (center right), and J0036+18 (bottom). The lighter overlaid lines show the model spectra of varying temperatures and gravities. The features in the data closely follow the features seen in the model spectra, indicating the absence of any H$_{3}^{+}$ emission features (see Section~\ref{sec:results}). The spectrum for DENIS~1058-15, shows some systematic differences at shorter wavelengths, possibly due to spectrophotometric variability or bad weather.}
	\label{fig:modelzoom1}
\end{figure*}

\begin{figure*}[htbp]
	\centering
	\includegraphics{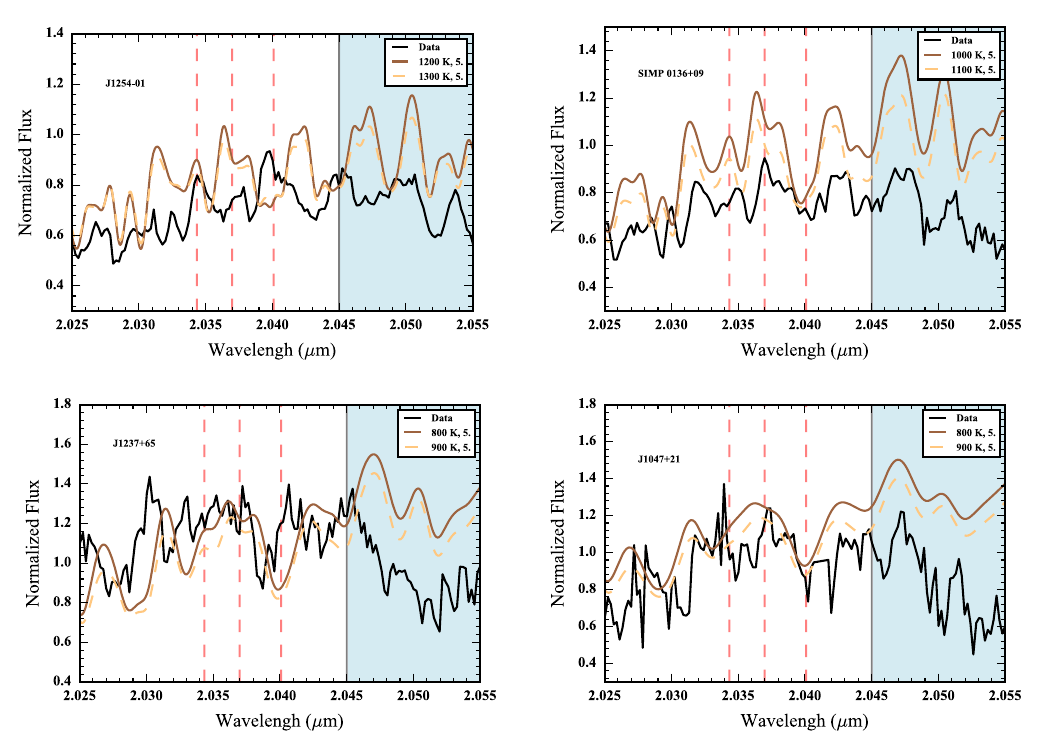} 
	\caption{ Same as Figure~\ref{fig:modelzoom1}, but for J1254-01 (top left), SIMP0136+09 (top right), J1237+65 (bottom left), and J1047+21 (bottom right). The strongest H$_{3}^{+}$ features are near a telluric band (shaded), but do not reveal anything prominently different than shape and features seen in model spectra.}
	\label{fig:modelzoom2}
\end{figure*}

To examine the spectra in detail, we compared the observations to BT-Settl PHOENIX model spectra convolved to the resolution of the data and zoomed in on the region with the strongest expected H$_{3}^{+}$ line \citep{Baraffe2015}. These plots are shown in Figures~\ref{fig:modelzoom1}-\ref{fig:modelzoom2}. In each plot, we show the data as a black solid line and the models as lighter solid and dashed lines. Vertical dashed lines mark off the same H$_{3}^{+}$ features from Figures~\ref{fig:MOScomp1}-\ref{fig:MOScomp2}. The models chosen are those from the model grids that best represent the atmospheric properties as determined in the literature (see Appendix~\ref{sec:targ}).

For LSR 1835 and TVLM 513 (top of Figure~\ref{fig:modelzoom1}), the model spectra match the data very well, each of the spectroscopic features can be matched clearly to corresponding features in the model. Although there is a slight excess in emission near the location of the H$_{3}^{+}$ feature at 2.093 $\mu$m, we do not see corresponding excesses at the locations of the other features in line with the relative ratios of the different emission line strengths. Moreover, a comparison between these two similar temperature M dwarfs, reveals the same features in both atmospheres, suggesting that neither shows evidence for H$_{3}^{+}$. Literature samples also show that this feature, relative to that at 2.091~$\mu$m, may be sensitive to temperature across the M-dwarf sequence \citep{RojasAyala2012ApJ...748...93R}.

For J0746+20AB (middle left of Figure~\ref{fig:modelzoom1}), we used a representative model atmosphere of T$_{\mathit{eff}}$ = 2100 K for the combined light spectrum of this nearly equal temperature binary system, consistent with an L1 combined light spectral type. Like the two previous targets, the features in the spectrum closely match those seen in the model with similar deviations as noted for LSR~1835 and TVLM~513, likely photospheric. 

Our plot of the comparison between our spectrum for DENIS~1058-15 and the model (middle right of Figure~\ref{fig:modelzoom1}), shows a broad difference in the K-band spectral shape. Because observing conditions for this target were not ideal, there were intermittent clouds and we were not observing at the parallactic angle, this difference is likely systematic. However, this target is also known to be photometrically variable in the NIR, potentially due to brown dwarf clouds and these could potentially cause spectroscopic variations in the emergent flux \citep[e.g.,][]{Heinze2013,Buenzli2014}. Nevertheless the absorption features in the spectrum are clear and are consistent with the features evident in the model spectrum.

Our spectrum of J0036+18 (botton of Figure~\ref{fig:modelzoom1}) is similar to those of LSR 1835 and TVLM 513. The spectral features are well distinguished and are in line with the model spectra. The same feature near 2.093 $\mu$m is apparent and observed with a similar shape as the other late M dwarfs and early L dwarfs (the nearest sky emission line is at 2.0909~$\mu$m). H$_{3}^{+}$ emission fluxes consistent with this excess are not seen at their expected locations. J0036+18 shows no indication of H$_{3}^{+}$ emission features. 

For the cooler T dwarfs in our sample, J1254-01, SIMP 0136+09, J1237+65, and J1047+20, the expected emission levels of H$_{3}^{+}$ based on typical H$\alpha$ emission strengths should be easily detected in our moderate resolution spectra. In Figure~\ref{fig:modelzoom2}, we plot the data as compared to the model PHOENIX spectra. These plots focus on the region around 2.04 $\mu$m where the most prominent H$_{3}^{+}$ lines are for temperatures below $\sim$1300 K. In each of these spectra, we do not see any indication of H$_{3}^{+}$ emission features. 

While it is plausible that individual objects may have variable emission level strengths and contrasts with the photospheric background, or across the rotationally variable surfaces aurorally impacted regions may not always have been visible, these concerns are unlikely to impact the observations of all 9 targets. Rather in corroboration of the results from \citet{Gibbs2022AJ....164...63G}, the brown dwarfs either have no H$_{3}^{+}$ emissions, or they are much weaker than anticipated. In the next Section, we consider the circumstances perhaps responsible for these observational results.

\begin{figure}[htbp]
   \centering
   \includegraphics[width=0.5\textwidth]{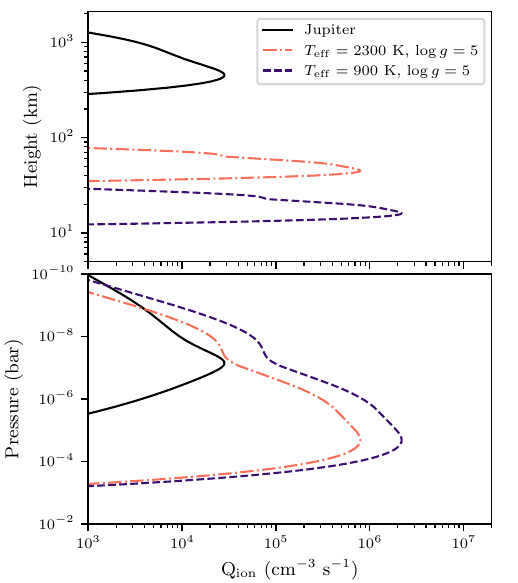} 
   \caption{Our model ion production rate for the same electron beam energy spectrum applied to Jupiter and brown dwarf atmospheric profiles (a late M-dwarf, 2300~K, and a T-dwarf, 900~K) give different peak H$_{3}^{+}$ production locations; See Section~\ref{sec:beamheight}. The two panels show the ion production rates at different points in the atmosphere according pressure (\textit{bottom}) and physical height above the 1 bar level (\textit{top}). The Jovian translation of height to pressure is based on the data from Galileo probe \citep{Seiff1997Sci...276..102S}.} 
   \label{fig:bdheight}
\end{figure}

\section{Electron Beam Interaction on Brown Dwarfs}\label{sec:ebeam}

If auroral brown dwarfs do not display H$_{3}^{+ }$ emissions, then their departure from expectations based on the Jovian system likely resides in differences in the interaction of the auroral precipitation with the brown dwarf atmosphere. Such differences with respect to Jupiter also appear evident in the UV \citep[e.g.,][]{Saur2021A&A...655A..75S}. In this Section, we explore how an auroral electron beam may interact with the brown dwarf, and the implications for the possibility of H$_{3}^{+ }$ emissions. We consider two questions: 1) Where in the brown atmosphere do we predominantly expect the auroral electron beam interaction? 2) Is there an appreciable population of H$_{3}^{+}$ available to generate IR auroral emission features?

\subsection{Height of Energy Deposition}\label{sec:beamheight}

\begin{figure*}[htbp]
   \centering
   \includegraphics[width=\textwidth]{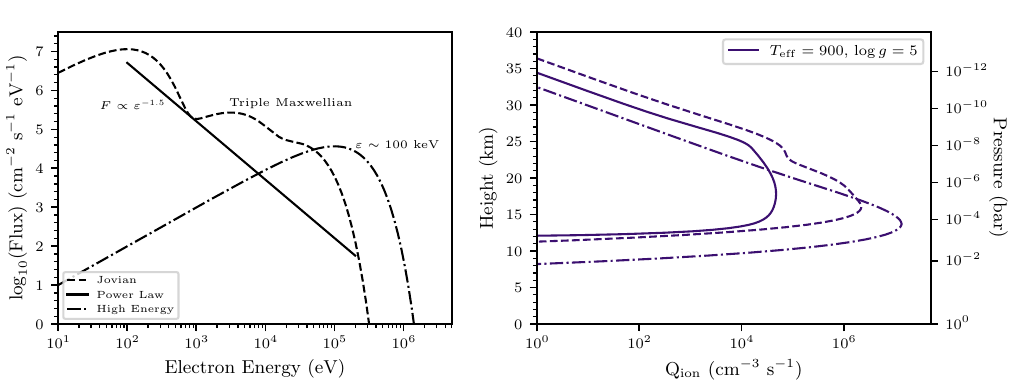} 
   \caption{\textit{Left} - Distinct electron beam spectra (Jovian-like, dashed; power-law, solid; and high-energy, dot-dashed) used to drive the ion production of H$_{3}^{+}$ in our model interaction with a brown dwarf atmosphere. \textit{Right} - Ion production rate profiles in example T-dwarf atmosphere in response to electron beam spectra at left.}
   \label{fig:beamspec}
\end{figure*}

To address our first question, we adapt the work of \citet{Hiraki2008AnGeo..26...77H} which parameterize Monte Carlo simulations for Jovian electron beam precipitation. They determined the rate of ionization from energetic electron impacts in a molecular hydrogen dominated atmosphere, including energy degradation, to define the generation of H$_{2}^{+}$ as a function of height in the atmosphere from a vertically propagating electron beam. Their simulation includes 10 different collisional processes (e.g., electronic, vibrational, and rotational excitation, ionization, etc.) that can both directionally scatter and/or remove energy from the electron beam. The fundamental simplifying assumptions are that the electron path can be traced through the bulk motion of the cyclotron guiding center, while ignoring collisions with any species besides molecular hydrogen. The latter assumption has only a $\sim$10\% effect on the height of peak ionization, and hence the localized production rate of H$_{3}^{+}$. The cyclotron motion may influence the collision rate of the weakest energy electrons in their simulations (1~eV), which continue to elastically scatter until they lose energy through rotational/vibrational excitation of H$_{2}$. We do not expect the guiding center assumption to alter the ionization profile, as the low energy electrons are generally below the ionization threshold, and higher energy electron motion is still captured well by this simplification.

The result of \citep{Hiraki2008AnGeo..26...77H} are verified with Jovian conditions, and their analytic parameterization provides a useful means of applying their results toward any molecular hydrogen dominated atmosphere, like those of brown dwarfs. The function, $q_{\mathrm{ion} }(\varepsilon, Z)$ (their equation~5), gives the ionization atmospheric profile for a given electron energy, $\varepsilon$. We can then integrate over an assumed electron beam energy flux, $F_{\varepsilon}$, to determine the atmospheric ionization rate, $Q_{\mathrm{ion} }$, throughout the atmosphere,

\begin{equation}
    Q_{\mathrm{ion}}(Z) = \int  q_{\mathrm{ion} }(\varepsilon, Z) F_{\varepsilon}(\varepsilon) \; d\varepsilon \, ,
\end{equation}

\noindent where $Q_{\mathrm{ion}}$ is in units of cm$^{-3}$ s$^{-1}$, and the height, $Z$, is relative to the 1 bar level. Because H$_{2}^{+}$ should quickly react to form H$_{3}^{+}$, this rate effectively gives the production rate of H$_{3}^{+}$. 

The ion production rate depends on the atmosphere and the assumed electron spectrum. We show in Figure~\ref{fig:bdheight} how the same electron beam energy spectrum produces different ion production rates between Jupiter \citep[density profile as in ][]{Hiraki2008AnGeo..26...77H} and brown dwarfs, as a function of height above the 1 bar pressure level. For the brown dwarf atmospheres, we used the cloudless Sonora-Bobcat models in radiative- convective equilibrium \citep{Marley2021ApJ...920...85M}, with $\log g = 5$, and effective temperatues of 900~K and 2300~K. We had to extend the model temperature-pressure profiles isothermally in hydrostatic equilibrium in order to determine the atmospheric conditions at low-pressures ($< 10^{-4}$ bar). This has the effect of fixing the volume mixing ratios of each species to the level at $10^{-4}$ bar for lower pressures. The result is consistent with the expectation that atmospheric mixing is more efficient at low pressures \citep{Mukherjee2022ApJ...938..107M}. Although it might not be totally accurate for isolated brown dwarfs, and other effects may influence the profile, this provides a reasonable first approximation to understand the effects of electron precipitation. Moreover, our ionization model depends almost exclusively on the H$_{2}$ density structure, dictated predominantly by the dwarf gravity and effective temperature.

Despite the changes in height at peak ionization for the different atmospheres (Figure~\ref{fig:bdheight}), the profile peaks actually take place at similar pressure levels between the two example brown dwarfs. In general, the physical extent of the ionization is determined by the atmospheric scale height, which is much smaller for a higher gravity brown dwarf relative to Jupiter, and smallest for the cooler T-dwarf. The corresponding height of peak ion production above the 1 bar level is thus lower in the brown dwarf regime. For Figure~\ref{fig:bdheight}, we employed the Triple Maxwellian electron energy distribution utilized by \citet{Grodent2001}, as representative of the Jovian electron beam spectrum. This function combines three distributions of the form

\begin{equation}
    F_{\varepsilon}( \varepsilon; \varepsilon_{0}  ) = A \frac{\varepsilon}{\varepsilon_{0}} e^{ - \varepsilon / \varepsilon_{0}}  \, ,
    \label{eq:Espec}
\end{equation}

\noindent where $A$ sets the overall beam current density in units of $e^{-}$ s$^{-1}$ cm$^{-2}$ eV$^{-1}$, and the characteristic energy, $\varepsilon_{0}$ defines the mode of the distribution. We demonstrate the impact of using different electron beam spectra on the ion production in Figure~\ref{fig:beamspec}, including comparisons against a power-law distribution, and a single high-energy beam with distribution according to Equation~\ref{eq:Espec}. The normalization of the electron flux scales the total ion production ($Q_{\mathrm{ion}}$), but is generally unknown. As such the absolute values of $Q_{\mathrm{ion}}$ may not be physically accurate, but the relative profile with height in the atmospheres (or pressure) is. In Figure~\ref{fig:beamspec}, we show that with broader electron spectra, we would expect a broader profile of ionization, with higher energy electrons penetrating more deeply into the atmosphere.

\begin{figure}[htbp]
   \centering
   \includegraphics[width=0.5\textwidth]{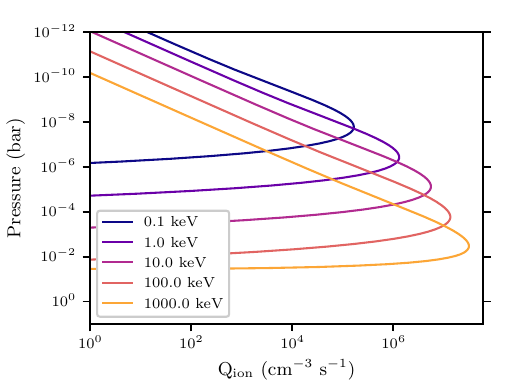} 
   \caption{Ionization rate profiles for the electron beam spectra following Equation~\ref{eq:Espec}, labeled according to their characteristic energies, $\varepsilon_{0}$, show peak ionization at deeper pressure levels with increasing energy, and a steep drop in ionization rate away from that peak. The example profiles here utilize the 900~K brown dwarf model, but are very similar for the 2300~K model (see Figure~\ref{fig:peakion}).}
   \label{fig:energpress}
\end{figure}

\begin{figure}[htbp]
   \centering
   \includegraphics[width=0.5\textwidth]{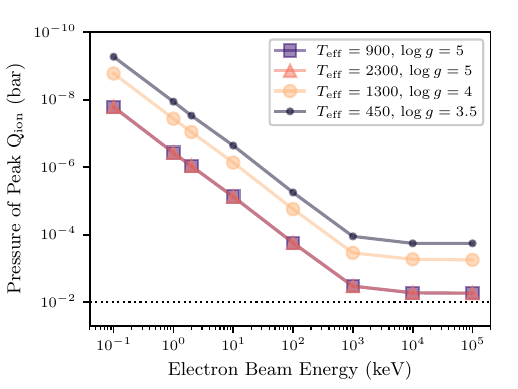 } 
   \caption{The pressure at which the peak ionization takes place for different characteristic electron energies goes deeper into the atmosphere until it appears to level off at high-energies. For the model brown dwarf atmospheres, the strongest beams appear to reach pressure levels of about $10^{-2}$~bar in $\log g = 5$ atmospheres. At lower gravity the height of peak ionization shifts higher up in the atmosphere to lower pressure for a fixed electron energy.}
   \label{fig:peakion}
\end{figure}

We illustrate this clearly in Figure~\ref{fig:energpress} with electron beam spectra utilizing a range of characteristic energies (Equation~\ref{eq:Espec}). Higher energy electrons generally also produce more ionization because after their secondary, tertiary, etc.\ impacts, the electrons still retain enough energy to continue to collisionally ionize H$_{2}$. We further illustrate the dependence of the pressure location (height in the atmosphere) of the peak ion production rate with characteristic electron beam energy ($\varepsilon_{0}$, using spectra as Equation~\ref{eq:Espec}) in Figure~\ref{fig:peakion}. The location of the ion production steadily declines as the stopping height goes lower into the atmosphere until it appears to level off for the highest energies ($\gtrsim 1$ MeV).

As a function of pressure, the location of the ionization peak is determined by the atmospheric gravity: similar gravity means similar pressures at peak ionization. Within the formalism of \citet{Hiraki2008AnGeo..26...77H}, the ionization rate is directly proportional to the mass density profile, but the profile shape is dictated by the depth to which individual electrons penetrate into the atmosphere. The latter is controlled by the column density (inversely proportional to gravity at fixed pressure). This is because, along the path traversed by the energetic electrons, the product of the column density with the interaction cross section sets the probability of ionizing collisions.

\citet{Hiraki2008AnGeo..26...77H} did not test their parameterization at the highest energies, they only went as high as $\sim200$~keV, but we illustrate the general idea here that higher energy beams go deeper into the atmosphere to some limit. Their results likely extend toward 1~MeV energies, but likely not higher. Furthermore, the atomic data at the foundation of the \citet{Hiraki2008AnGeo..26...77H} simulations have not been experimentally verified at high energies ($>1$~keV), although more recent collision cross section studies remain consistent with the data used by those authors \citep[e.g.,][]{Karwasz2022MolPh.12070087K}. Nevertheless, their results are consistent with observational studies of the Jovian ionization profiles, lending confidence to their work, and our extension of it. These results for pressure at peak ionization are also consistent with the work of \citet{Helling2019RSPTA.37780398H}, who modeled electron beam interactions with brown dwarf atmospheres, including transport effects, but with limited electron energies. Full electron ion impact models on brown dwarfs need to verify the beam stopping heights toward the highest electron energies. Our work and similar studies would benefit greatly by experiments and theory verifying electron impact cross sections in higher energy collisions.

Higher energy electron beams could be expected as a consequence of magnetospheric-ionospheric coupling in brown dwarfs with stronger magnetic fields and more rapid rotation, relative to Jupiter. Strong effective field aligned voltages (equivalent to $\sim0.5$~MeV) are required to support the current generated in the brown dwarf electrodynamic engines \citep{Turnpenney2017MNRAS.470.4274T}. These electron beam energies in the range from 5~keV to 1~MeV, are still permissible for ECM generation \citep{Dulk1985ARA&A..23..169D}. Additionally, high energy electrons may also be needed to model the radio beam properties of the detected GHz emissions \citep{Lynch2015}. We do not expect auroral electron beams to exist beyond 1~MeV energies. The electron cyclotron maser instability is limited to only mildly relativistic particles (higher energies lead to gyrosynchrotron and synchrotron emissions) \citep{Dulk1985ARA&A..23..169D, Treumann2006A&ARv..13..229T}. Electron beam spectra much more energetic than the Jovian case, reaching several hundred keV energies, are consistent with our available observational constraints on auroral brown dwarfs.

Detections of atmospheric auroral emissions, and the determination of where that energy is being deposited should provide a constraint on the typical electron beam energies. Auroral energy deposition at deep levels ($10^{-2}$ bar) may indicate energetic beams of up to 1~MeV, with deposition predominantly at higher altitudes giving indications of weaker electron beam energies.

\begin{figure*}[htbp]
   \centering
   \includegraphics[width=0.49\textwidth]{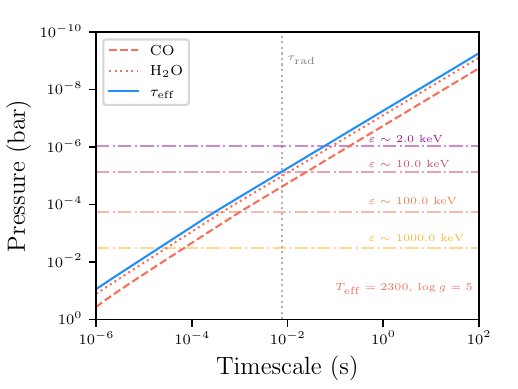} 
   \includegraphics[width=0.49\textwidth]{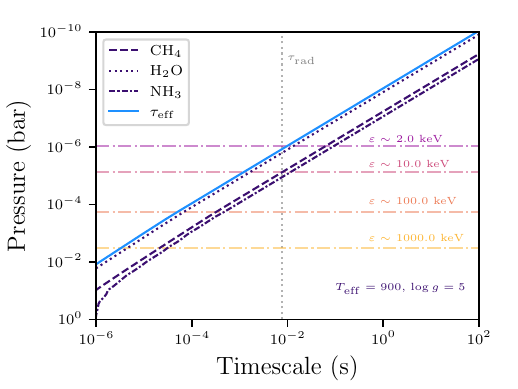}
   \caption{ For the late M-dwarf (\textit{left}) and T-dwarf (\textit{right}) model atmospheres, we show the corresponding timescale of dominant ion chemistry pathways to destruction of H$_{3}^{+}$, including the total effective timescale (solid blue).  Deeper in the atmosphere the destruction proceeds more quickly than the radiative de-excitation timescale (vertical dotted line), i.e., the curves are to the left of the vertical line. The depth of peak ionization and H$_{3}^{+}$ production for different energy electron beams are indicated with horizontal lines. }
   \label{fig:timescales}
\end{figure*}

\subsection{Lifetimes and Emission Time Scales}

In order for H$_{3}^{+}$ to be detected as molecular emission it must survive long enough in the atmosphere, once created, in order to radiatively de-excite. The results of the previous subsection underscore how the sustained presence of H$_{3}^{+}$ depends on the typical electron beam energies, because deeper in the brown dwarf atmosphere the molecular ion is increasingly likely to react with molecules that would destroy H$_{3}^{+}$ --- an idea first presented in \citet{Pineda2017}.

H$_{3}^{+}$ is very reactive and will be destroyed in brown dwarf atmospheres through reactions of the form

\begin{equation}
    \mathrm{H}_{3}^{+} + \mathrm{X} \rightarrow \mathrm{HX}^{+} + \mathrm{H}_{2} \; .
    \label{eq:h3pX}
\end{equation}

\noindent Most any typical species can take on the role of $\mathrm{X}$ in brown dwarf atmospheres, including CH$_{4}$, CO, and H$_{2}$O. For each such reaction we can define a time scale associated with the rate at which a particular species is destroying H$_{3}^{+}$,

\begin{equation}
    \tau_{a} = [k_{a} n_{a}]^{-1} \; , 
\end{equation}

\noindent where $k_{a}$ is the reaction rate, typically in units of cm$^{3}$ s$^{-1}$. For faster rates (greater $k_{a}$) and a greater number density of the particular species ($n_{a}$), then the time scale is smaller, i.e., quicker. These timescales do not depend on the amount of H$_{3}^{+}$, but instead on the atmospheric properties wherever there is any H$_{3}^{+}$.

Comparing the time scales of the dominant reactions driving the destruction of H$_{3}^{+}$ with the radiative time scale for IR emissions lets us determine whether there is enough time for the molecule to appreciably emit before being destroyed. Since multiple species allow for multiple pathways to decrease the population of H$_{3}^{+}$, we define the effective destruction timescale as

\begin{equation}
    \tau_{\mathrm{eff} } \equiv \left[ \sum \tau^{-1}_{a} \right]^{-1} \; ,
\end{equation}

\noindent where the sum is over all relevant species interacting with H$_{3}^{+}$ as in Equation~\ref{eq:h3pX}. The timescale changes throughout the atmosphere as the density of each species changes. We show the comparison of these timescales in Figure~\ref{fig:timescales}, using the model example atmospheres of the late M-dwarf (\textit{left}), and T-dwarf (\textit{right}), and the known reaction rates taken from the STAND2019 chemical network \citep{Rimmer2016ApJS..224....9R,Rimmer2019Icar..329..124R}. In Figure~\ref{fig:timescales}, for each model atmosphere and at each pressure level, we show the corresponding destruction timescale for each species (dashed lines, etc.), and the total effective timescale (solid blue line). The radiative time scale (inverse of the Einstein A coefficient) for the strongest H$_{3}^{+}$ IR emission lines is $\sim10^{-2}$~s \citep{Neale1996,Tao2011}.  Deeper in the atmosphere at higher pressures, the destruction timescales are smaller than the radiative timescale, indicating that at those heights H$_{3}^{+}$ does not survive long enough to radiatively de-excite, and produce the IR emission line features. For the brown dwarfs these pressures correspond to energy deposition and ionization heights for electron beams with energies $\gtrsim2-10$~keV. Only beams with low energy electrons (similar to Jovian and weaker) are likely to generate any H$_{3}^{+}$ from the lowest pressure regions of the atmosphere. At these low pressure regions ($< 10^{-6}$-$10^{-7}$ bar) the mixing of the molecular species have likely decoupled \citep{Woitke2020A&A...634A..23W}, above the homopause, so destruction timescales are actually longer than modeled in this study, favoring H$_{3}^{+}$ emission at those heights.

If the beams are very energetic, as anticipated by the modeling of the brown dwarf magnetospheric-ionospheric coupling currents \citep{Nichols2012ApJ...760...59N,Turnpenney2017MNRAS.470.4274T}, then brown dwarf atmospheres are unlikely to exhibit any H$_{3}^{+}$ emissions. This atmospheric chemistry would explain this work and literature observations showing no evidence for any IR H$_{3}^{+}$ emission lines. Our analysis further suggests a rough lower limit on typical late M-dwarf and T-dwarf electron beam energies of 10~keV, and 2~keV respectively, with beams needing less than this energy typically to generate long-lasting levels of H$_{3}^{+}$ in the brown dwarf upper atmospheres. Well defined luminosity limits for the H$_{3}^{+}$ lines would map directly to upper limits on the electron beam flux, $F_{\varepsilon}$, at low-energies; however, a full model of the ionization and emissions in varied substellar atmospheres will be required to take this next step \citep[e.g.,][]{Tao2011,Tao2012}.

\section{Discussion}\label{sec:discuss}

Our results (Section~\ref{sec:results}) indicate that the brown dwarfs with known signatures of auroral emissions, predominantly radio, do not emit appreciable amounts of H$_{3}^{+}$ in the K-band, in line with the results of \citet{Gibbs2022AJ....164...63G} at L' band from a smaller sample of brown dwarfs. While it is possible that our medium resolution data (R$\sim$3600) may not have been able to distinguish the emission lines, especially if they were weak, the higher resolution data of \citet{Gibbs2022AJ....164...63G} would not have suffered from that issue (R$\sim$25,000). Since the radio pulses and optical variability imply the existence of an energetic electron beam interacting with the atmosphere, we take this as a given and consider the physical scenarios that limit the emission of H$_{3}^{+}$ features. 

Our assumed auroral energy scaling was based on the auroral emissions of Jupiter. 
However, important differences in the atmospheric properties and potentially the auroral electron energy distribution may lead to diminished H$_{3}^{+}$ emissions. 
Of particular concern could be the atmospheric temperature structure and/or the prevalence of species destroying H$_{3}^{+}$.

The amount of H$_{3}^{+}$ emission is sensitive to the atmospheric temperature on both Jupiter and Saturn, as seen in auroral atmospheric models, with emissions generally increasing with higher temperatures and the Kronian emissions being more sensitive to temperature than the Jovian ones \citep{Tao2011}. However, auroral modeling has not yet considered the higher atmospheric temperatures expected in the brown dwarfs, and it is unclear, although perhaps unlikely, if the effect continues to temperatures as high as $\sim$2000 K. Nevertheless, we can comment on the influence of temperature on the T~dwarf emissions. Despite upper atmospheres that have temperatures that are not likely too much higher than Jovian exosphere temperatures (T$\sim $1000 K), we do not see any emission in the T dwarfs \citep[e.g.,][]{Lam1997,Grodent2001}. Consequently, the difference in temperatures is not likely to drive the differences in H$_{3}^{+}$ emission.

Non-LTE effects suppress the strength of emission features, especially at high latitudes where the densities are low \citep{Melin2005,Tao2011}. Our model spectrum assumed LTE conditions, and is likely systematically biased, however our rough estimates were conservative in nature and the effect of the non-LTE conditions is only a factor of a few on the assumed total energy \citep{Tao2011}. Despite this caveat, there should still have been significant H$_{3}^{+}$ emission, if the processes were identical to the Jovian case. 

Through electron recombination, a large electron number density will destroy any H$_{3}^{+}$ that is produced. In the absence of heavy molecules like H$_{2}$O or CH$_{4}$, this recombination process is the dominant loss mechanism for H$_{3}^{+}$ \citep{Badman2015}. Although the brown dwarfs are warmer and can sustain more thermally ionized atmospheres than what is expected for giant planets, the atmospheres are still predominantly neutral and are not likely to suppress H$_{3}^{+}$ entirely through electron recombination \citep{RodriguezBarrera2015}.

Although the H$_{3}^{+}$ ion emits readily on time scales $\sim$0.01 s, high density environments can lead to rapid reactions that remove H$_{3}^{+}$ before it has an opportunity to emit \citep{Badman2015}. In Saturn's atmosphere, H$_{2}$O plays this role, whereas in Jupiter, CH$_{4}$ is the dominant reacting species \citep{Perry1999,Tao2011,Badman2015}. We explored this effect for brown dwarfs (Section~\ref{sec:ebeam}), comparing the destruction timescales for H$_{3}^{+}$ at the layers where the electron beam impact ionization rates peak. We found that for beams with typical electron energies only factors of several stronger than the Jovian case, the deposition layers are deep enough that reactions with H$_{2}$O prevent appreciable emissions of H$_{3}^{+}$ in both the late M-dwarf and T-dwarf atmospheres. Furthermore, this result is robust to the introduction of significant atmospheric mixing, since the volume mixing ratio of H$_{2}$O is not significantly different throughout the atmosphere at the temperatures from late M-dwarfs to T-dwarfs, although the relative contributions of destruction via CO or CH$_{4}$ will change \citep[e.g.,][]{Mukherjee2022ApJ...938..107M}.

The observed absence of H$_{3}^{+}$ emission potentially indicates that the brown dwarf electron energy distributions have sufficient energy to penetrate into these deeper layers. Our calculations suggest that electron beam energies of at least $2$-$10$~keV are required for T-dwarfs, and late M-dwarfs respectively to explain the non-detections. Higher electron beam energies relative to the Jovian case are consistent with current modeling of the magnetospheric-ionospheric coupling of brown dwarf auroral currents \citep{Nichols2012ApJ...760...59N,Turnpenney2017MNRAS.470.4274T}. These current properties are observationally constrained by the strong radio aurorae seen throughout the sample population studied in this work \citep[e.g.,][]{Williams2015a, Hallinan2015}. This is a consequence of the faster rotation of the brown dwarfs, a few hours relative to the 10 hr Jovian rotation period, and the stronger magnetic field strengths, kG fields instead of 10 G fields. Broad electron energy spectra with appreciable fluxes of lower energy electrons may still contribute to the production of H$_{3}^{+}$ at low pressures; however, limits on the IR line emissions should place upper limits on this lower energy electron flux, but requires detailed modeling of the production and emission of H$_{3}^{+}$, including atmospheric effects, like heat transport.

If the brown dwarf electron beams are more energetic, and the energy is predominately deposited at deeper layers there are likely additional consequential differences from the Jovian scenario. The destruction of H$_{3}^{+}$ via H$_{2}$O may present an opportunity to use hydronium features as auroral indicators, as also mentioned by \citet{Helling2019RSPTA.37780398H}. A deep penetrating electron beam would lead to diminished ultraviolet aurorae, H$_{2}$ Werner bands, due both to collisional de-excitation, and a significant hydrocarbon absorption in the column of gas above the auroral energy deposition layer \citep{Gustin2013}, consistent with existing non-detections \citep{Saur2018ApJ...859...74S, Saur2021A&A...655A..75S}. A high collision rate at the electron energy deposition layer should quickly thermalize the energetic electrons and significantly heat the atmosphere. Without emission features like H$_{3}^{+}$ to strongly cool the upper atmosphere, the bulk auroral energy would emerge almost entirely as thermal energy through the aurorally modified atmospheric thermal profile. The lack of UV and IR auroral detections do not suggest an absence of electron beam precipitation, but rather that auroral emissions must be considered within the unique context of their host atmospheres and magnetospheres.

These multi-wavelength considerations and the observational evidence reinforce the idea that brown dwarf atmospheres significantly depart from the Jovian auroral energy partition \citep{Saur2018ApJ...859...74S}. With few alternative mechanisms for losing the additional energy deposited through electron beam precipitation, the total thermal heat may account for almost the entirety of the ultimate release of the auroral energy. Any additional brown dwarf auroral emission features are mostly a consequence of the resulting atmospheric temperature inversion (H$\alpha$). 

In the future, fully self-consistent atmospheric modeling of the auroral impact on brown dwarfs are needed to elucidate the detailed physical constraints and define new observational tests; however, our approach has delineated a consistent framework across existing brown dwarf auroral observations from UV, optical, IR, and radio studies requiring high energy auroral electron beams. The H$_{3}^{+}$ non-detections point specifically to the electron beam energy minima (or weak flux of low energy electrons), and additional work on the thermal structure of auroral objects should provide the concrete measurements needed to further constrain the electrodynamic engines powering brown dwarf aurorae. Extrasolar searches for aurorally generated H$_{3}^{+}$ due to electron beams will benefit from a focus on targets with lower gravity (see Appendix~\ref{sec:appEx}), and/or weaker electrodynamic engines (weaker magnetic fields, slower rotation etc.); under these conditions, the production of H$_{3}^{+}$ is much more likely to generate detectable emission line features.

\section*{Acknowledgments}

The authors wish to recognize and acknowledge the very significant cultural role and reverence that the summit of Mauna Kea has always had within the indigenous Hawaiian community.  We are most fortunate to have the opportunity to conduct observations from this mountain.

We thank the anonymous reviewer for their thoughtful comments in improving this manuscript. J.S.P would like to thank Chuck Steidel and Gwen Rudie for their assistance preparing observations with MOSFIRE, as well as Nick Konidaris for help with the data reduction. J.S.P also thanks Michael Eastwood for observing support, and Caroline Morley \& James Mang for useful model discussions. 

J.S.P was supported by a grant from the National Science Foundation Graduate Research Fellowship under grant No. (DGE-11444469). J.S.P acknowledges some support for this work under program HST-15924, and JWST-1874 provided by NASA through grant from Space Telescope Science Institute, operated by the Association of Universities for Research in Astronomy, incorporated under NSA contract NAS5-26555.

The data presented herein were obtained at the W.M. Keck Observatory, which is operated as a scientific partnership among the California Institute of Technology, the University of California and the National Aeronautics and Space Administration. The Observatory was made possible by the generous financial support of the W.M. Keck Foundation.

This publication makes use of data products from the Two Micron All Sky Survey, which is a joint project of the University of Massachusetts and the Infrared Processing and Analysis Center/California Institute of Technology, funded by the National Aeronautics and Space Administration and the National Science Foundation.

\begin{figure*}[htbp]
   \centering
   \includegraphics[width=\textwidth]{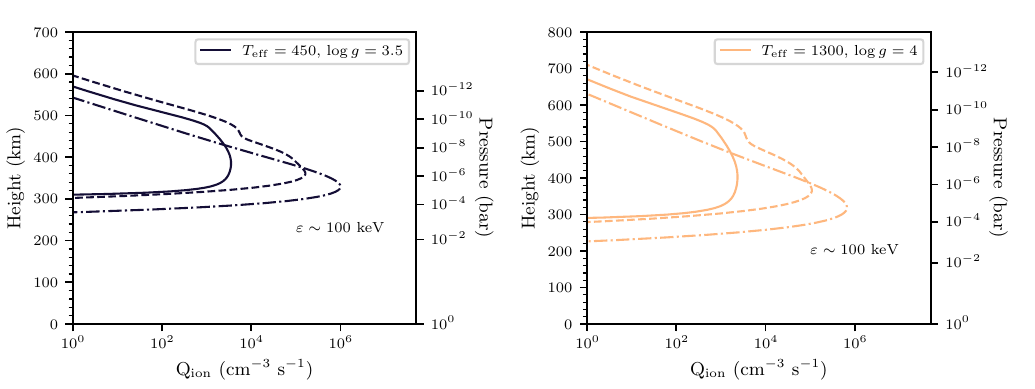} 
   \caption{ Applying the electron beam spectra shown in Figure~\ref{fig:beamspec}, we show the corresponding ionization rate profiles for example model atmospheres with properties similar to Y-dwarfs (\textit{left}), and to young planetary mass companions (\textit{right}). }
   \label{fig:appendixProf}
\end{figure*}

\appendix

\section{Additional Example Electron Beam Impacts}\label{sec:appEx}

For the known auroral population of brown dwarfs, spanning late M-dwarfs to late T-dwarfs, the example model atmospheres discussed in Section~\ref{sec:ebeam} focused on low and high temperature examples with similar gravity ($\log g = 5$). Our analysis further showed that the atmospheric gravity dictates the pressure level of peak ionization for electrons of a given energy (see Figure~\ref{fig:peakion}). Other cool objects, not yet confirmed to display auroral behaviour, may also possess auroral electron beams. 

For interested readers we apply the analyses of Section~\ref{sec:ebeam} to two additional model atmospheres: 1) a Y-dwarf-like atmosphere, of very cool effective temperature (T$_{\mathit{eff}} = $~450~K) and low gravity ($\log g = 3.5$), and 2) a warm late L dwarf atmosphere (T$_{\mathit{eff}} =$ 1300~K) of low gravity ($\log g = 4$), similar to the properties of the known population of young planetary mass companions \citep[e.g.,][]{Bowler2020AJ....159...63B}. In Figure~\ref{fig:appendixProf}, we show the ionization profiles through these atmospheres when assuming the electron energy spectra shown in the left panel of Figure~\ref{fig:beamspec}. For the lower gravity atmospheres, the ionization peak (proxy for energy deposition) shifts to lower pressure, and higher up in the atmosphere. 

\begin{figure*}[htbp]
   \centering
   \includegraphics[width=0.49\textwidth]{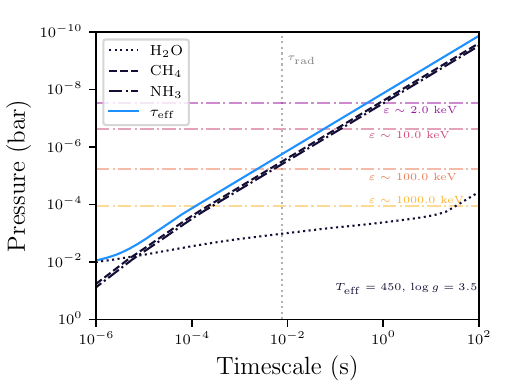}
   \includegraphics[width=0.49\textwidth]{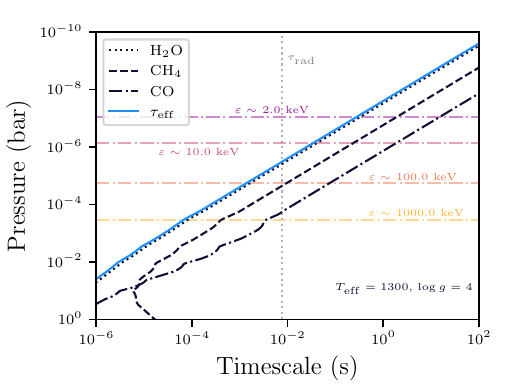}
   \caption{Same as Figure~\ref{fig:timescales}, but applied to the two additional example atmospheres, a Y-dwarf (\textit{left}; T$_{\mathit{eff}} = 450$ K, $\log g = 3.5$), and a young planetary mass companion (\textit{right}; T$_{\mathit{eff}} = 1300$ K, $\log g = 4$).   }
   \label{fig:appendixtime}
\end{figure*}

Considering the destruction timescales of H$_{3}^{+}$, as in Section~\ref{sec:scale}, we further show the main species hypothetically limiting H$_{3}^{+}$ in these model atmospheres in Figure~\ref{fig:appendixtime}. For the young planetary mass object (\textit{right} panel), water remains the main species driving the destruction of H$_{3}^{+}$ throughout the atmosphere. Whereas, for the Y-dwarf (\textit{left} panel), the atmosphere is so cold that at low pressures the water has rained out, and methane is the dominant species destroying H$_{3}^{+}$. At these lower gravities, stronger beams are required to go deeper into the atmosphere, where the reaction timescales are shorter than the emission timescale of H$_{3}^{+}$. Lower gravity atmospheres where energy deposition and ionization take place higher up are better targets in search of aurorally generated H$_{3}^{+}$ emissions.

\section{Target Details}\label{sec:targ}

We selected a sample of UCD targets with indications of potential auroral activity, usually radio emission and H$\alpha$, but also photometric variability, to look for H$_{3}^{+}$ emission. These targets are summarized in Table~\ref{tab:obslog}; we discuss each in detail here. We also include here literature reports of effective temperature and gravity to be used in a comparison between our spectra and model atmospheres (See Section~\ref{sec:results}).

\subsection*{LSR J1835+3259}

\noindent Discovered by \cite{Reid2003}, LSR~J1835+3259 (hereafter LSR~1835), a M8.5 dwarf with a K-band magnitude of 9.171 at a distance of $5.67$ pc, has been studied extensively for its magnetic activity \citep{Skrutskie2006}. LSR~1835 is known to emit sinusoidally varying H$\alpha$ emission and displays steady long-term photometric variability with a period of 2.84 hrs \citep{Harding2013, Hallinan2015}. LSR~1835 also shows both quiescent radio emission indicative of a strong radiation belt \citep{Kao2023Natur.619..272K}, and periodically pulsed highly polarized radio emission, spanning 4 - 8 GHz \citep{Hallinan2008, Hallinan2015}. Despite the presence of strong magnetic fields, as indicated by the radio emission, no X-ray detections have been observed from LSR~1835+32 \citep{Berger2008,Berger2010}. \cite{Berger2008} also report near ultraviolet ($\lambda_{\mathrm{eff}}$ = 2600~\AA) measurements from \textit{Swift} consistent with photospheric emission. \cite{Filippazzo2015} report an effective temperature, T$_{\mathit{eff}}$, of 2316~$\pm$~51 K and gravity, $\log$ g, of 5.22 $\pm$ 0.11 (cgs) base on a semi-empirical approach using evolutionary models and measured bolometric luminosities from multi-wavelength observations spanning the bulk of the spectral energy distribution of the target. \citep{Gibbs2022AJ....164...63G} report emission limits of $<6.3$-$6.7\times10^{16}$~W.

\subsection*{TVLM513-46546}

\noindent TVLM513-46546 (hereafter TVLM 513) is a M8.5 dwarf with a K-band magnitude of 10.706 at a distance of 10.59 pc \citep{Kirkpatrick1995,Dahn2002,Skrutskie2006}. TVLM~513 has been studied extensively for its magnetic activity, including both H$\alpha$ and radio emission and has become a benchmark object for activity at the end of the main sequence. TVLM~513 displays highly polarized periodic pulses at a rotation period of 1.958 hr and quiescent radio emission \citep{Hallinan2007, Hallinan2008}. Extensive follow-up at radio wavelengths has shown this emission and period to be very stable and long-lasting \citep{Wolszczan2014}. TVLM~513 also shows stable photometric variations over long time periods at optical wavelengths, that could be connected to the process emitting the radio emission \citep{Harding2013,Wolszczan2014}. TVLM~513 has an effective temperature and log gravity of 2242~$\pm$~55 K and 5.22~$\pm$~0.11 (cgs), respectively \citep{Filippazzo2015}.

\subsection*{2MASS~J07464256+2000321}

\noindent 2MASS~J07464256+2000321 (hereafter J0746+20AB) is a binary system comprising L0 and L1.5 dwarfs at a distance of 12.2 pc with a K-band magnitude of 10.468 \citep{Dahn2002,Bouy2004, Skrutskie2006}. Extensive mulit-wavelength monitoring of this target has revealed periodic radio pulses and H$\alpha$ emission that are offset in phase by 0.25 within a period of 2.07 hr, as well as non-detections in the X-ray and UV \citep{Berger2009}. Photometric monitoring of the binary by \cite{Harding2013b} indicates that the primary component is photometrically variable at optical wavelengths and rotates with a period of 3.3 hr, whereas the faster rotating secondary is the source of the radio emission.

\subsection*{DENIS J1058.7-1548}

\noindent Discovered by \cite{Delfosse1997}, DENIS~J1058.7-1548 (hereafter DENIS~1058-15) is a L3 dwarf at a distance of 17.3 pc with a K-band magnitude of 12.532 \citep{Kirkpatrick1999,Dahn2002,Skrutskie2006}. Spectroscopic studies detected weak H$\alpha$ emission from DENIS~1058-15 \cite[e.g.,][]{Kirkpatrick1999}. This target also displays J-band variability in photometric monitoring with similar periodic variation in the \textit{Spitzer} bands at 3.6 and 4 $\mu$m \citep{Heinze2013,Metchev2015}. DENIS~1058-15 has an effective temperature and log gravity of 1809~$\pm$~68 K and 5.20~$\pm$~0.19 (cgs), respectively \citep{Filippazzo2015}.

\subsection*{2MASS J00361617+1821104}

\noindent 2MASS~J00361617+1821104 (hereafter J0036+18) is an L3.5 dwarf at a distance of 8.76 pc with a K-band magnitude of 11.058 \citep{Reid2000,Kirkpatrick2000,Dahn2002,Skrutskie2006}. Initial studies demonstrated that J0036+18 displayed strong radio emission and periodic highly circularly polarized pulses \citep{Berger2002,Berger2005,Hallinan2008}. However, simultaneous monitoring campaigns showed no X-ray emission and an apparent lack of H$\alpha$ \citep{Berger2005,Berger2010}. \cite{Pineda2016a}, however, did detect H$\alpha$ emission, indicating variability timescales much longer than the rotation period of 3.08 hr \citep{Hallinan2008}. \cite{Harding2013} detected periodic variability, consistent with the rotational period in optical photometric monitoring. Additionally, \cite{Metchev2015} detected irregularly periodic photometric variability in \textit{Spitzer} monitoring of J0036+18 at both 3.6 and 4 $\mu$m. J0036+18 has T$_{\mathit{eff}}$ = 1869~$\pm$~64 K and $\log$ g = 5.21~$\pm$~0.17 (cgs; \citealt{Filippazzo2015}).

\subsection*{2MASS J12545393-0122474}

\noindent Discovered by \cite{Leggett2000}, 2MASS~J12545393-0122474 (hereafter J1254-01) is a brown dwarf with an infrared spectral type of T2 at a distance of 13.21 pc with a K-band magnitude of 13.84 \citep{Vrba2004,Skrutskie2006,Burgasser2006}. \cite{Burgasser2003} report the detection of weak H$\alpha$ emission from J1254-01. Further monitoring at radio wavelengths yielded no detections in two 2 hr observing blocks \citep{Kao2016}. \textit{Spitzer} observations looking for photometric variability at 3.6 and 4 $\mu$m also did not yield positive results \citep{Metchev2015}. J1254-01 has T$_{\mathit{eff}}$ = 1219~$\pm$~94 K and $\log$ g = 5.02~$\pm$~0.47 (cgs; \citealt{Filippazzo2015}).

\subsection*{SIMP J013656.5+093347.3}

\noindent SIMP~J013656.5+093347.3 (hereafter SIMP~0136+09), discovered by \cite{Artigau2006}, is a T dwarf at the L/T transition, with a K-band magnitude of 12.562, displaying strong photometric variability in the J-band interpreted as heterogenous cloud cover in its atmosphere \citep{Skrutskie2006, Artigau2009}. At a distance of 6.1 pc, SIMP~136+09 is one of the closest cool brown dwarfs and a benchmark object for atmospheric modeling \citep{Weinberger2016}. Efforts to find H$\alpha$ emission have only produced upper limits despite multiple observations at different rotational phases \citep{Pineda2016a}. However, radio monitoring at 4 GHz by \cite{Kao2016} has detected quiescent emission and multiple highly circularly polarized radio pulses. \cite{Kao2016} use an updated version of the spectral index methods of \cite{Burgasser2006b} to measure T$_{\mathit{eff}}$ = 1089$\pm^{62}_{54}$ K and $\log$ g = 4.79$\pm^{0.26}_{0.33}$ (cgs) for SIMP~0136+09. Kinematic analysis of this target has also shown that it is a likely member af the Carina-Near moving group, has an age less than 950 Myrs, and an estimated mass of 12 M$_{\mathrm{Jup}}$ \citep{Gagne2017}.

\subsection*{2MASS J12373919+6526148}

\noindent Since its discovery, 2MASS~J12373919+6526148 (hereafter J1237+65), a brown dwarf at 10.4 pc with a K-band magnitude of 16.4, has been an unusual T6.5 (infrared spectral type) because of its exceedingly strong H$\alpha$ emission \citep{Burgasser1999,Burgasser2003,Vrba2004,Burgasser2006,Skrutskie2006}. Many follow-up efforts have attempted to understand the cause of the emission, however the initial studies were inconclusive \cite[][]{Burgasser2002, Liebert2007}. More recently, the search for ECM radio emission in cool brown dwarfs has revealed J1237+65 to be a radio source with moderate levels of circular polarization, likely connected to the strong H$\alpha$ emission \citep{Kao2016}. J1237+65 has T$_{\mathit{eff}}$ = 851~$\pm$~74 K and $\log$ g = 4.95~$\pm$~0.5 (cgs; \citealt{Filippazzo2015}).

\subsection*{2MASS J10475385+2124234}

First reported in \cite{Burgasser1999}, 2MASS~J10475385+2124234 (J1047+21 hereafter) is a brown dwarf with a spectral type of T6.5 in the infrared, at a distance of $10.56$ pc, with a K-band magnitude of 16.2 \citep{Vrba2004,Burgasser2006,Skrutskie2006}. \cite{Burgasser2003} detected weak H$\alpha$ emission from this target, as one a few T dwarfs with this emission feature. J1047+21 was also the first T dwarf to be detected in radio emission \citep{Route2012}. Follow-up efforts have since confirmed the radio detection, measured quiescent emission levels and characterized the periodic pulses of the object with a period of 1.77 hr \citep{Williams2015a, Kao2016, Route2016}. J1047+21 has an effective temperature and log gravity of 880~$\pm$~76 K and 4.96~$\pm$~0.49 (cgs), respectively \citep{Filippazzo2015}.

\bibliography{h3p}
\bibliographystyle{aasjournal}

\end{document}